\documentclass[aps,pre,onecolumn,showpacs,showkeys,amsmath,amssymb,
groupedaddress]{revtex4}
\usepackage{graphicx}

\newcommand{\prt}{\partial}

\begin{document}

%\preprint{}

\title{A dynamical systems approach to a thin accretion disc and
its time-dependent behaviour on large length scales}

\author{Arnab K. Ray}
\email{arnab@mri.ernet.in}
\affiliation{Harish--Chandra Research Institute \\
Chhatnag Road, Jhunsi, Allahabad 211019, INDIA}

\author{J. K. Bhattacharjee}
\email{tpjkb@mahendra.iacs.res.in}
\affiliation{Department of Theoretical Physics \\
Indian Association for the Cultivation of Science \\
Jadavpur, Kolkata 700032, INDIA}

\date{\today}

\begin{abstract}
Modelling the flow in a thin accretion disc like a dynamical system, 
we analyse the nature of the critical points of the steady solutions 
of the flow. For the simple inviscid disc there are two critical points, 
with the outer one being a saddle point and the inner one a centre 
type point. For the weakly viscous disc, there are four 
possible critical points, 
of which the outermost is a saddle point, while the next inner one is 
a spiral. Coupling the nature of the critical points with the outer 
boundary condition of the flow, gives us a complete understanding of 
all the important physical features of the flow solutions in the 
subsonic regions of the disc. In the inviscid disc, the physical 
realisability of the transonic solution passing through the 
saddle point is addressed by considering a 
temporal evolution of the flow, which is a very likely non-perturbative
mechanism for selecting the transonic inflow solution from among a 
host of other possible stable solutions. 
For the weakly viscous disc, while a linearised time-dependent 
perturbation imposed on the steady mass inflow rate causes instability, 
the same perturbative analysis reveals that for the inviscid disc, 
there is a very close correspondence between the equation for the 
propagation of the perturbation and the metric of an acoustic black 
hole. Compatible with the transport of 
angular momentum to the outer regions of the disc, a viscosity-limited 
length scale is defined for the spatial extent of the inward rotational 
drift of matter. 
\end{abstract}

\pacs{47.40.Hg, 05.45.-a, 98.62.Mw}

\keywords{Transonic flows, Nonlinear dynamical systems, Accretion and 
accretion disks}

\maketitle

\section{Introduction}

In astrophysical fluid dynamics, studies in accretion processes
occupy a position of prominence. Of such processes, critical (transonic)
flows --- flows which are regular through a critical point --- are of
great importance~\cite{skc90}. A classic example of such a flow is the 
Bondi solution in steady spherically symmetric 
accretion~\cite{bon52,skc90}. What is striking about the Bondi solution 
is that while the question of its realisability is susceptible to great 
instabilities arising from infinitesimal deviations from an absolutely 
precise prescription of the boundary condition in the steady limit of 
the hydrodynamic flow, it is easy to lock on to the Bondi solution when 
the temporal evolution of the flow is followed. This dynamic and 
non-perturbative selection mechanism of the transonic solution agrees 
quite closely with Bondi's conjecture that it is the 
criterion of minimum energy that will make the transonic solution 
the favoured one~\cite{bon52,gar79}. 

The appeal of the spherically symmetric flow, however, is limited by 
its not accounting for the fact that in a realistic
situation, the infalling matter would be in possession of angular
momentum --- and hence the process of infall should lead to the formation
of what is known as an accretion disc. A simple and well-understood 
model of such an accreting system is the inviscid and thin accretion 
disc~\cite{az81,yk95,msc96,wit99,dpm03}, 
and this we will initially consider in our study to
provide ourselves with an effective guideline for a subsequent approaching
of the more general viscous problem. However, the inviscid model has 
its own limitations because, while the presence of angular momentum 
leads to the formation of an accretion disc, a 
physical mechanism must be found for the outward transport of the 
angular momentum, which should then make possible the infall of the
accreting matter into the potential well of the accretor. Viscosity
has been known to be such a physical means to effect infall, although
the exact presciption for viscosity in an accretion disc is still 
a matter of much debate~\cite{fkr92}. What is well appreciated, however, 
is that the viscous prescription should be compatible with an enhanced 
outward transport of angular momentum. The very well known $\alpha$ 
parametrisation of Shakura and Sunyaev~\cite{ss73, fkr92} is based on 
this principle, and we invoke this $\alpha$ model to logically extend our 
study from an inviscid disc to a ``quasi-viscous" disc. With that 
objective we have prescribed a first-order viscous perturbation 
(dependent on $\alpha$) about the zeroth-order 
inviscid solution, and we see that the merest 
presence of viscosity in the flow equations, gives rise to significant
qualitative changes in the flow behaviour, insight about which could not 
have been derived from the inviscid disc model alone. 

In our present study, the accretion flow has been considered to be driven
by the classical Newtonian potential. For accretion on to an ordinary
star or even a white dwarf, this is arguably a satisfactory working
premise. We set up the governing steady flow equations of both the 
inviscid and the quasi-viscous disc like a dynamical system. This 
approach gives us a clear understanding of the nature of the critical 
points of the flow, as well as any sense of direction that may be 
associated with each solution. Combining this understanding with a 
knowledge of the outer boundary condition for physically meaningful 
inflow solutions, makes it possible to construct the phase diagram of 
the flow solutions for both the cases we are studying. 
For the inviscid disc it has been found that there are exactly two
critical points for the steady flow solutions, of which, for realistic
boundary conditions, the outer one is a saddle point, while the inner
one is a centre type point. The situation gets somewhat more complicated
for the quasi-viscous disc. In this second case we see that there should
be four critical points, of which we analyse the behaviour of those 
two which, in the limit of vanishing $\alpha$ would take us back to 
the phase portrait of the inviscid flow. The outer one of these two 
critical points remains saddle type as before, while
the inner one behaves like a stable spiral.
While dwelling on this matter, we should mention that studies on the
nature of the critical points in an accretion disc have been reported
before. While numerically studying the effect of $\alpha$-viscosity
on the flow structure around the inner edge of the disc near a 
non-rotating black hole, Matsumoto et al.~\cite{mat84} discussed the 
behaviour of critical points for the phase portrait of the isothermal 
flow, and for small $\alpha$ they upheld the case for a transonic saddle
point. Along similar lines Afshordi and Paczy\'nski~\cite{ap03} adopted 
a semi-analytic approach to understanding the nature of the fixed 
points for constructing a physical solution. In making a quantitative 
analysis of the inner boundary condition, for a
constant effective speed of sound, they also found the existence
of a saddle point for small $\alpha$. 
While mentioning these background issues, we must stress however,
that our interest in the disc would be to study its stability aspects
and its long-time evolutionary properties on large length scales, and
to do so we would also need to have a clear notion of at least the
qualitative features of the phase portrait of the steady flow solutions,
and its critical points. That these important 
and qualitatively physical conclusions about the flow could be drawn
without taking any recourse to the conventional practice of a 
numerical integration of the steady flow equations, amply
demonstrates the simplicity, the elegance and the power of the 
dynamical systems approach that we have adopted to study the thin 
accretion disc. 

Following our understanding of the nature of the critical points in
the phase portrait, the question that we then take up is about the 
preference of the 
accreting system for any particular velocity profile in the stationary
phase portrait, and a selection criterion thereof. In this regard
the transonic solution has always been the favoured candidate. However, 
it is common knowledge from the study of dynamical systems, that a
flow solution (the transonic flow in our case) passing through a saddle 
point cannot be realised physically~\cite{js77}. To address this issue 
satisfactorily it must be 
appreciated that the real physical flow is not static in nature, but 
has an explicit time-dependence. With respect to this point it is 
tempting to subject the steady flow solutions to small perturbations 
in real time, and then study their behaviour. This has been done 
elsewhere~\cite{ray03} for the inviscid disc, and it has been shown 
that the steady inflow solutions of abiding interest are all stable 
under the influence of a linearised time-dependent perturbation on the 
mass inflow rate. 
The application of a similar treatment on the quasi-viscous disc,
however, very much differently affects the stability of its steady 
inflow solutions, which, as we have shown here, become destabilised by 
small perturbations --- treated both as standing and travelling waves. 

Since one way or the other, no direct conclusion could be drawn about the 
selection of a particular solution through a perturbative technique,
for the inviscid disc at least, we try to have an understanding of a 
true selection mechanism and the attendant choice of a 
particular solution, by studying the evolution of the accreting 
system through real time. A model analog shows that it is indeed possible 
in principle for the temporal evolution to allow for the selection 
of an inflow solution that passes through the saddle point, and under
restricted conditions we demonstrate that the selection criterion 
conforms to Bondi's minimum energy argument, which is invoked to favour 
the transonic solution in the spherically symmetric case. Interestingly
enough in this context, we have also shown that although the perturbative 
study has offered no direct clue about the selection of a solution, the 
equation governing the propagation of an acoustic disturbance in the 
flow bears a close resemblance with the effective metric of an acoustic
black hole. We make use of this similarity to argue that the flow would
cross the acoustic horizon transonically. 

For the quasi-viscous disc, dissipation upsets all idealised conditions,
and allows for the setting of no precise criterion for selecting any
particular solution, something that we might 
otherwise establish under conserved conditions. Nevertheless, we have
at least been able to argue that with the cumulative transfer of angular 
momentum to the outer regions of the disc, a characteristic scale of 
length may be set, beyond which the accumulation of angular momentum 
resists further inward drift of matter. 

\section{The steady equations of the flow and its critical points}

The flow that we consider is an axisymmetric thin disc whose local 
height is given by $H$~\cite{fkr92}. It is customary to describe the
steady flow by the equation of continuity, the equation for radial
momentum balance and the equation for angular momentum balance, all of
which are respectively given in the following notation~\cite{ny94} by
\begin{equation}
\label{con}
\frac{\mathrm d}{\mathrm{d}R}\left(\rho vRH \right)=0
\end{equation}
\begin{equation}
\label{radmom}
v{\frac{\mathrm{d}v}{\mathrm{d}R}} + \frac{1}{\rho} 
\frac{\mathrm{d}P}{\mathrm{d}R} + 
\left(\Omega_{\mathrm{K}}^2 -\Omega^2 \right)R = 0
\end{equation}
\begin{equation}
\label{angmom}
v \frac{\mathrm d}{\mathrm{d}R}\left(\Omega R^2 \right) =
\frac{1}{\rho RH}\frac{\mathrm d}{\mathrm{d}R} \left (
\frac{\alpha \rho c_{\mathrm s}^2 R^3 H}{\Omega_{\mathrm K}} 
\frac{\mathrm{d} \Omega}{\mathrm{d}R} \right )
\end{equation}
In the above, $v$ is the radial drift velocity, $R$ is the radial 
distance, $\Omega$ is the angular velocity, $\Omega_{\mathrm K}$ is the 
Keplerian angular velocity defined as $\Omega_{\mathrm K}^2=GM/R^3$, 
$c_{\mathrm s}$ is the velocity of sound defined as 
$c_{\mathrm s}^2={\partial}P/{\partial}{\rho}$ and $\alpha$ is the 
effective viscosity parameter of Shakura and Sunyaev~\cite{fkr92, ny94}. 

It is a standard practice to make use of a general polytropic equation 
of state $P=k{\rho}^{\gamma}$ where $k$ and $\gamma$ are constants, 
with $\gamma$ being the polytropic exponent, whose admissible range 
$(1< \gamma < 5/3)$ is restricted by the isothermal limit and the 
adiabatic limit, respectively. The condition of hydrostatic equilibrium 
along a direction perpendicular to the thin disc, allows us to use the 
approximation~\cite{fkr92} 
\begin{equation}
\label{thinapprox}
\frac{H}{R}{\cong}\frac{c_{\mathrm s}}{v_{\mathrm K}}
\end{equation} 
where $v_{\mathrm K}=R \Omega_{\mathrm K}$. In that case Eq.(\ref{con}) 
leads to 
\begin{equation}
\label{coninteg}
c_{\mathrm s}^{2n+1}vR^{5/2}= \mathrm{constant}
\end{equation}
in which $n=(\gamma -1)^{-1}$. The constant of integration in 
Eq.(\ref{coninteg}) can be physically identified with the mass inflow 
rate. Combining Eqs.(\ref{radmom}) and (\ref{coninteg}) gives,
\begin{equation}
\label{dvdR}
\frac{\mathrm{d}\left(v^2 \right)}{\mathrm{d}R} = \frac{2v^2}{R}
\left [ \frac{5 c_{\mathrm s}^2 - \left( \gamma +1 \right)
\left (\Omega_{\mathrm K}^2- \Omega^2 \right)R^2}
{\left( \gamma +1 \right) v^2-2 c_{\mathrm s}^2} \right ]
\end{equation}
and a first integral of Eq.(\ref{angmom}) can be written as 
\begin{equation}
\label{anginteg}
\alpha \frac{\mathrm d}{\mathrm{d}R}\left(\Omega^2\right)=2 \Omega 
\left ( \Omega -\frac{L}{R^2} \right ) \frac{v \Omega_{\mathrm K}}
{c_{\mathrm s}^2}
\end{equation}
in which $L$ is a constant of integration.

The inviscid limit of Eq.(\ref{anginteg}) may be obtained by setting
$\alpha =0$. This model has found some use in accretion 
studies~\cite{az81,msc96}, and it gives us
the condition $\Omega R^2=L$, where $L$, which can be physically
identified as the specific angular momentum of the flow, now becomes 
a constant of the motion. Such a constraint gives us an explicit 
dependence of $\Omega$ on $R$, and allows us to fix the critical points 
of the flow in the $v^2$ --- $R$ space. At such points both the numerator 
and the denominator in the right hand side of Eq.(\ref{dvdR}) vanish 
simultaneously~\cite{skc90}, and this will deliver the critical
point conditions as 
\begin{align}
\label{critcon}
v_{\mathrm c}^2 &=  \frac{2 c_{\mathrm{sc}}^2}{\gamma +1}  \nonumber \\
\frac{5 c_{\mathrm{sc}}^2}{\gamma +1} &= \left(\Omega_{\mathrm{Kc}}^2
-\Omega_{\mathrm c}^2\right) R_{\mathrm c}^2
\end{align}
with the subscripted label $\mathrm{c}$ indicating critical point values. 
The first condition gives the flow velocity at the critical point in 
terms of the local speed of sound at the same point. The latter 
condition, when written down explicitly as 
\begin{equation}
\label{rcdep}
\frac{5 c_{\mathrm{sc}}^2}{\gamma +1} = \frac{GM}{R_{\mathrm c}} - 
\frac{L^2}{R_{\mathrm c}^2}
\end{equation}
gives a dependence of $R_{\mathrm c}$ on $c_{\mathrm{sc}}$. This is a 
quadratic in $R_{\mathrm c}$, indicating that there are two critical points. 
The critical points $(v_{\mathrm c}^2, R_{\mathrm c})$ could be fixed in 
the $v^2$ --- $R$ space, if $c_{\mathrm{sc}}$ could be expressed in terms 
of the constants of the flow system. To do so, it should be necessary for 
us to look at the integral of Eq.(\ref{radmom}), which reads as
\begin{equation}
\label{radinteg}
\frac{v^2}{2}+ \frac{L^2}{2R^2}- \frac{GM}{R} + \frac{c_{\mathrm s}^2}
{\gamma -1}= E
\end{equation}
in which the constant of integration $E$, which is actually the 
Bernoulli constant, can be determined with the help of the boundary 
condition that for very great radial distances, the speed of sound 
approaches a constant ``ambient" value $c_{\mathrm s}(\infty)$, while 
the radial drift velocity $v$ becomes vanishingly small. On using the 
critical point conditions in Eq.(\ref{radinteg}), it could be seen
that $c_{\mathrm{sc}}$ could in principle be represented solely in 
terms of the physical constants of the system. Futhermore, for this
flow system, we require that $c_{\mathrm{s}}$ should be a monotonically 
increasing function for decreasing $R$, except perhaps very close to 
the accretor~\cite{fkr92}, and therefore $c_{\mathrm{s}}$ should be 
single-valued at all $R$. On physical grounds alone, this should be 
a perfectly valid expectation. This would imply that for every value 
of $R_{\mathrm c}$, there should be a single physical value of 
$c_{\mathrm{sc}}$, given by the profile of $c_{\mathrm{s}}$, and
this value of $c_{\mathrm{sc}}$, having been fixed in terms of the
system constants, would consequently enable the critical points to 
be fixed as well in the phase space of the flow solutions. This 
conclusion may be alternatively and equivalently derived by fixing 
$c_{\mathrm{sc}}$ with the help of Eq.(\ref{coninteg}). 

Thus the two solutions of the quadratic in $R_{\mathrm c}$, as given 
by Eq.(\ref{rcdep}), are obtained as 
\begin{equation}
\label{critpoints}
R_{\mathrm c}=\frac{\gamma +1}{10}\frac{GM}{c_{\mathrm{sc}}^2}
\left [ 1 \pm \sqrt{1 - \frac{20}{\gamma +1}
\left ( \frac{c_{\mathrm{sc}}L}{GM}\right )^2} \right ]
\end{equation}
and the two fixed points may be considered to be at 
radii $R=R_{\mathrm{c1}}$ and $R=R_{\mathrm{c2}}$, with 
$R_{\mathrm{c2}}>R_{\mathrm{c1}}$. We note from Eq.(\ref{critpoints}) 
that real values of $R_{\mathrm c}$ would necessitate the condition that 
for a regular solution passing through a critical point, there would be 
an upper bound to the admissible value of the constant angular momentum $L$. 

So far we have used the inviscid solution $\Omega = LR^{-2}$, derived
by requiring that $\alpha =0$. About this inviscid solution for small
non-zero values of $\alpha$, we now prescribe a ``quasi-viscous" solution,
whose form, to a first order in $\alpha$, we may write as 
\begin{equation}
\label{quasol}
\Omega = \frac{L}{R^2} + \alpha \tilde{\Omega}
\end{equation}
in which the unknown function $\tilde{\Omega}$ is to be determined by 
invoking the
angular momentum balance condition. To that end we use Eq.(\ref{quasol}) 
in Eq.(\ref{anginteg}), and by dropping all orders in $\alpha$ higher
than the first, we will get 
\begin{equation}
\label{formofl}
\tilde{\Omega} = - \frac{2L c_{\mathrm s}^2}{v v_{\mathrm K} R^2}
\end{equation}
which will allow us to define an effective angular momentum as 
\begin{equation}
\label{effangmom}
L_{\mathrm{eff}}(R) = R^2 \Omega = L \left( 1 -   
\frac{2 \alpha c_{\mathrm s}^2}{v v_{\mathrm K}} \right)
\end{equation}
Using this new functional dependence of $\Omega$ in the radial momentum
balance condition, given by Eq.(\ref{radmom}), we can account for
the manner in which outward angular momentum transport effects the 
inward drift of matter. The validity of this dependence may be put to
the test by studying the asymptotic relation of the angular momentum
drift. On very large length scales the drift velocity $v$ goes 
asymptotically as $R^{-5/2}$. Bearing in mind that for inflows, $v$ is 
negative, on scales of length where the fluid assumes ambient conditions,
we will have 
\begin{equation}
\label{asympangmom}
L_{\mathrm{eff}} \sim L + 2\alpha L 
\left(\frac{R}{R_L} \right)^3
\end{equation} 
Here $R_L$ is a suitable scale of length, which, to an 
order-of-magnitude is given by 
$R_L^3 \sim GM \dot{m}[c_{\mathrm s}^3(\infty) 
\rho (\infty)]^{-1}$, with $\dot{m}$ being the mass inflow rate. This 
asymptotic behaviour is what we should expect entirely, because the 
physical role of viscosity is to transport angular momentum to large 
length scales of the accretion disc. 

Going back now to the critical point conditions, given by 
Eq.(\ref{critcon}), we see that while the form of the first condition
remains unaltered, the second condition, under the quasi-viscous
regime (i.e. first order in $\alpha$), gives 
\begin{equation}
\label{visrcdep}
\frac{5 c_{\mathrm{sc}}^2}{\gamma +1} = \frac{GM}{R_{\mathrm c}} -
\frac{L^2}{R_{\mathrm c}^2} + 2 \alpha \left(\gamma +1 \right)
\frac{L^2}{R_{\mathrm c}^2} \left ( \frac{v_{\mathrm c}} 
{v_\mathrm{Kc}} \right )
\end{equation}
in which $v_\mathrm{Kc} = \sqrt{GMR_{\mathrm c}^{-1}}$ is the Kepler 
velocity at the fixed points. We can therefore, eventually write  
Eq.(\ref{visrcdep}) as a fourth-degree equation in 
$R_{\mathrm c}$, provided, as before for the inviscid case, 
$c_{\mathrm{sc}}$ is fixed in
terms of the system constants. To do so, unlike before, we can only
turn to the continuity condition, given by Eq.(\ref{coninteg}),
since with viscous dissipation now present, the radial momentum 
balance equation cannot be exactly integrated. With $c_{\mathrm{sc}}$
thus fixed, we conclude that in this new quasi-viscous scenario,
there should be four critical points. Of these, we would be interested
in those two points which, in the limit of $\alpha \longrightarrow 0$,
would satisfactorily reproduce all the features of the inviscid flow. 

\section{The accretion disc as a dynamical system}

An integration of the first-order differential equation, given by
Eq.(\ref{dvdR}), for both the inviscid and the quasi-viscous disc,
should in principle give us all the possible flow solutions in the 
$v^2$ --- $R$ space, corresponding to every possible value of the 
integration constant (itself to be determined with the help of 
boundary conditions for physical flows). This, however, could only 
be done numerically because the speed of sound is a function of both 
the radial drift velocity and the radial distance. So far this has 
been the conventional method of studying this problem. We contend 
here that alternatively a complete understanding of the behaviour 
of the flow solutions could be had, if standard techniques of 
dynamical systems analysis could be introduced to study an inviscid 
and thin accretion disc. To do that --- having first determined the
number of critical points --- it would then be important for us to 
identify the nature of each of the critical points of the flow system. 
To that end we turn to Eq.(\ref{dvdR}) and parametrise it to get 
\begin{align}
\label{dyn}
\frac{\mathrm{d}(v^2)}{\mathrm{d} \tau} &= 2v^2 
\left [ \frac{5 c_{\mathrm s}^2}{\gamma +1} -
\left(\Omega_{\mathrm K}^2- \Omega^2 \right)R^2 \right ] \nonumber \\
\frac{\mathrm{d}R}{\mathrm{d} \tau} &= R \left [ v^2-
\frac{2 c_{\mathrm s}^2}{\gamma +1} \right ]
\end{align}

This parametrisation has been carried out in a mathematical parameter 
space $\tau$, and in such a space the two equations above represent 
an autonomous first-order dynamical system~\cite{js77}, given in the 
standard mathematical form ${\dot{y}}=Y(x,y)$ and ${\dot{x}}=X(x,y)$.
To analyse the nature of the fixed points in this parameter space, 
$v^2$ and $R$ would have to be expanded and then linearised in the
expanded terms in Eq.(\ref{dyn}), about the fixed point coordinates 
$(R_{\mathrm c},v_{\mathrm c}^2)$. 

For the inviscid disc, in terms of the expanded quantities $\delta v^2$ 
and $\delta R$, a set of linear equations will then be given by 
\begin{align}
\label{lindyn}
\frac{1}{2 v_{\mathrm c}^2}
\frac{\mathrm d}{\mathrm{d} \tau}\left(\delta v^2 \right) &=  
-{\frac{5}{2}} \left({\frac{\gamma -1}{\gamma +1}}\right)
{\delta}v^2+ \left [ \left ( \frac{6-4{\gamma}}{\gamma +1} \right ) 
\frac{GM}{R_{\mathrm c}^2} - \left (\frac{7-3 \gamma}{\gamma +1} \right ) 
\frac{L^2}{R_{\mathrm c}^3} \right ] \delta R  \nonumber \\
\frac{1}{R_{\mathrm c}}
\frac{\mathrm d}{\mathrm{d} \tau}\left(\delta R \right) &= 
\frac{2 \gamma}{\gamma +1} \, {\delta}v^2 + 
5 \left ( \frac{\gamma -1}{\gamma +1} \right ) 
\frac{v_{\mathrm c}^2}{R_{\mathrm c}} \, {\delta}R 
\end{align}

For the pair of linear first-order differential equations above, we use  
solutions of the form $e^{\lambda \tau}$ for the perturbed quantities 
and this would deliver the eigenvalues $\lambda$ --- growth rates of 
${\delta}v^2$ and ${\delta}R$ in $\tau$ space --- as
\begin{equation}
\label{eigen}
\lambda^2=4 \frac{5- \gamma}{\left(\gamma +1 \right)^2}
\frac{GM}{R_{\mathrm c}} c_{\mathrm{sc}}^2
\left[\zeta - \left( \frac{L}{L_\mathrm{Kc}}\right)^2 \right]
\end{equation}
where $\zeta = (5-3 \gamma)(5- \gamma)^{-1}$ and 
$L_\mathrm{Kc} = \sqrt{GMR_\mathrm{c}}$ is the local Keplerian
angular momentum at the fixed points.

\begin{figure}[t]
\begin{center}
\includegraphics[scale=0.5, angle=-90]{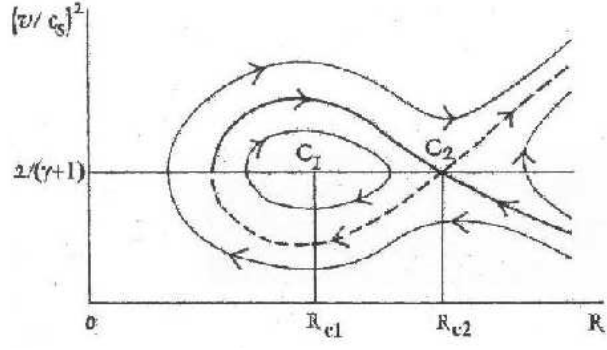}
\caption{\label{f.1} \small{A schematic diagram of the accretion problem.
The fixed point $\rm {C_2}$ is a saddle point. $\rm {C_1}$ is a centre
type point. The separatrices pass through $\rm {C_2}$.}}
\end{center}
\end{figure}

The term in the square brackets in Eq.(\ref{eigen}) will determine the
sign of $\lambda^2$. Knowing that two adjacent fixed points cannot 
be of the same nature~\cite{js77}, i.e. they will differ in their 
respective signs of $\lambda^2$, it is now evident that if $R_\mathrm{c}$ 
is the outer fixed point $R_\mathrm{c2}$, then $\lambda^2$ will be positive 
and the fixed point will be saddle type, while if $R_\mathrm{c}$ is the 
inner fixed point $R_\mathrm{c1}$, then $\lambda^2$ will be negative and 
the fixed point will be centre type. The flows of physical interest here 
are governed by the boundary condition that at large distances the drift 
velocity becomes vanishingly small,
while the speed of sound approaches a constant value. This knowledge of
the boundary condition, in conjunction with the nature of the fixed 
points, makes it possible to draw the phase trajectories of the flow,
which are as shown in Fig.~\ref{f.1}. The arrows in the figure, 
indicating the 
direction that may be associated with each flow solution, could only
be drawn by taking recourse to a dynamical systems analysis.

In a manner of speaking, the fixed point $\mathrm{C_2}$ at 
$R=R_\mathrm{c2}$,
which is a saddle point, may be dubbed the ``sonic point" because one of 
the two trajectories passing through it is a transonic inflow solution, 
rising from subsonic values far away from $\mathrm{C_2}$ to attain supersonic 
values for $R<R_\mathrm{c2}$. Here transonicity is to be attained when 
the drift
velocity equals the speed of the propagation of an acoustic disturbance,
which, for this system is given 
by $\sqrt{2}(\gamma +1)^{-1/2}c_\mathrm{s}$~\cite{ray03}. The 
transonic trajectory is shown as the heavy solid curve, and is an
accretion inflow solution. Once the flow is in the supersonic region,
the drift velocity starts decreasing for $R<R_\mathrm{c1}$. This  
happens because the centrifugal force in the flow (connected to its
angular momentum, which is not lost under inviscid conditions) builds 
up resistance against the attractive influence of gravity. 

We can now study how these idealised conditions are affected when 
viscosity is brought into play. This we can do by using 
Eq.(\ref{effangmom}) to substitute for $\Omega$ in Eq.(\ref{dyn}). In
carrying out a first-order perturbative expansion about the critical
point values, as we have done before for the inviscid case, we will 
get the relation 
\begin{align}
\label{vislindyn}
\frac{1}{2 v_{\mathrm c}^2} 
\frac{\mathrm d}{\mathrm{d} \tau}\left(\delta v^2 \right) 
& = \left [ - \frac{5}{2} \left( \frac{\gamma -1}{\gamma +1} \right)
+ \alpha \frac{\left(3 \gamma -1 \right) L^2}{v_{\mathrm c} R_{\mathrm c}
L_{\mathrm{Kc}}} \right ] {\delta}v^2 \nonumber \\
& + \left [ \left ( \frac{6-4{\gamma}}{\gamma +1} \right )
\frac{GM}{R_{\mathrm c}^2} - \left (\frac{7-3 \gamma}{\gamma +1} \right)
\frac{L^2}{R_{\mathrm c}^3} + \alpha \frac{2 \left ( 13\gamma -7 \right) L^2 
c_{\mathrm{sc}}^2}{\left(\gamma + 1 \right) v_{\mathrm c} R_{\mathrm c}^2 
L_{\mathrm{Kc}}} \right ]  \delta R  
\end{align}
The form of the second parametrised relation from Eq.(\ref{dyn}) 
remains unchanged
even with the introduction of viscosity, and this, together with 
Eq.(\ref{vislindyn}), will eventually deliver a quadratic equation for 
the eigenvalues that reads as 
\begin{equation}
\label{viseigen}
\lambda^2 - 2 \alpha \mathcal{X}_1 \lambda - \left(\mathcal{X}_2 
+ \alpha \mathcal{X}_3 \right) = 0
\end{equation}
where 
\begin{align}
\label{cees}
\mathcal{X}_1 & = \left(3 \gamma -1 \right) 
\frac{v_{\mathrm c}L^2}{R_{\mathrm c}
L_{\mathrm{Kc}}} \nonumber \\
\mathcal{X}_2 & = 4 \frac{5- \gamma}
{\left(\gamma +1 \right)^2} \frac{GM}{R_{\mathrm c}} c_{\mathrm{sc}}^2
\left[\zeta - \left ( \frac{L}{L_\mathrm{Kc}}\right)^2 \right] \nonumber \\
\mathcal{X}_3 & = 4 \left(\frac{11 \gamma -5}{\gamma +1}\right)
\frac{v_{\mathrm c}L^2}{R_{\mathrm c}L_{\mathrm{Kc}}} 
c_{\mathrm{sc}}^2
\end{align}

\begin{figure}[t]
\begin{center}
\includegraphics[scale=0.5, angle=-90]{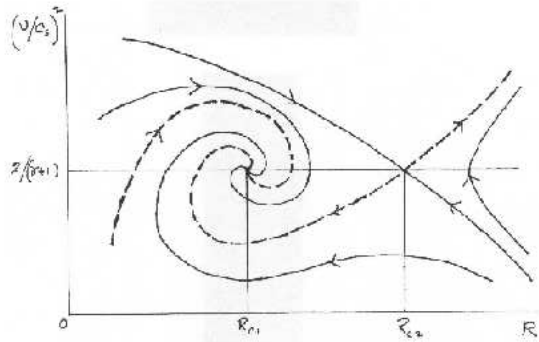}
\caption{\label{f.2} \small{A schematic diagram of the of the flow
solutions in a weakly viscous accretion disc. The outer fixed point
is a saddle point, but the next inner fixed point behaves like
a spiral in the presence of viscous dissipation.}}
\end{center}
\end{figure}

We can now solve the quadratic in $\lambda$ to write, 
\begin{equation}
\label{lambdasol}
\lambda = \alpha \mathcal{X}_1 \pm \sqrt{\alpha^2 \mathcal{X}_1^2 
+ \mathcal{X}_2 + \alpha \mathcal{X}_3}
\end{equation}
in which, the viscosity parameter $\alpha$ can be tuned to arbitrarily 
small (though not exactly zero) values. This should be consistent with 
our quasi-viscous approximation, and this will also make 
$\mathcal{X}_2$ the dominant term in the discriminant of
Eq.(\ref{lambdasol}). As before, we can make use of the fact that 
no two adjacent critical points can be of the same nature. Combining
this with the standard boundary condition, we can once again say that
the outermost critical point will be a saddle point. However, an 
interesting change appears in the next inner critical point. For 
$\alpha = 0$, we have seen that this point will be centre type, as 
has been shown in Fig.~\ref{f.1}. Now a look at Eq.(\ref{lambdasol})
immediately reveals that for very small non-zero values of $\alpha$,
with $\mathcal{X}_1$ being real and $\mathcal{X}_2 < 0$, this point
becomes a spiral. This behaviour has been diagrammatically represented
in Fig.~\ref{f.2}, and one can intuite that this is physically what it 
should be. With viscosity present, there is dissipation in the system,
and the solutions therefore spiral in, as opposed to the conserved
inviscid situation, where the solutions about the point $\mathrm{C_1}$
form closed paths. Another important point to note is that with 
dissipation being present in the system, there would have been no 
general means for the integration of the equation for momentum balance. 
In that event, this dynamical systems analysis affords an easy way
of studying the system.  

The dynamical systems approach, however, gives rise to a serious 
difficulty. The demonstration that the outermost fixed point is a 
saddle point, concomitantly places great difficulties in the way of 
realising the transonic trajectory passing through this point. This 
fact should be immediately evident from the direction that the arrows
(characteristic of a saddle point) indicate for the solutions passing 
through $\mathrm{C_2}$ in Fig.~\ref{f.1}. Conventional wisdom about the 
nature of saddle points leads to the understanding here that in the 
steady limit of the physical flows, the transonic solution would not 
be realisable~\cite{js77}. 
This issue may be understood in two ways. The first is 
more physical and is related to the prescription of the boundary 
condition. A look at Fig.~\ref{f.1} makes it quite obvious that 
the transonic solution could only be achieved after an infinitely 
precise determination of the outer boundary condition needed to 
generate it. Any deviation, however infinitesimally small, would 
take the system far away from transonicity. In the 
astrophysical context, it may easily be imagined that it is well nigh
impossible to have such precise fulfilment of the boundary condition
for transonicity. 

The second condition against achieving transonicity is somewhat more
mathematical in nature. To find out the eigenvalues of the set of linear 
equations given by Eq.(\ref{lindyn}), we have used a solution of the 
form $\delta R \sim e^{\lambda \tau}$ (the same form being used for
$\delta v^2$ as well), which we can recast as 
$\vert \tau \vert \sim {\lambda}^{-1} \ln \vert {\delta R}/{\tilde{R}} 
\vert$, in which $\tilde{R}$ is a suitable length scale factor to 
render the argument of the logarithm dimensionless. 
In the $\delta v^2$ --- $\delta R$ space, reaching the saddle point 
$(0,0)$, would then entail an infinitely great divergence of 
$\vert \tau \vert$. This shows that any attempt to numerically 
integrate Eq.(\ref{dvdR}) would be thwarted even if the practical 
impossibility of determining an absolutely precise outer boundary
condition were to be achieved. In Fig.~\ref{f.1} this fact is represented
by the way the arrows behave along the critical solutions, establishing
that they are not physical solutions per se, but are actually 
separatrices of various classes of solutions, and this
special status of the separatrices, as distinct from all other solutions, 
has been emphasised by the heavy solid curve and the heavy dashed curve
passing through $\mathrm{C_2}$. These arguments evidently apply for
the saddle point even in the presence of viscosity. 

However, it is presently a firmly established fact that transonic solutions
do occur in astrophysical systems such as the one being studied here. 
The difficulty about the realisability of the transonic solutions could
be suitably addressed by appreciating that the discussion presented so
far has been confined within the steady (time-independent) framework
of the hydrodynamic flow, and that the arrows representing the 
direction of the flows are in a mathematical parameter space. On
the other hand, in the real astrophysical context, the accreting system
has an explicit time-dependence, and is evolving through time. Taking 
this fact into account could therefore satisfactorily resolve all 
difficulties regarding attaining transonicity. 

In passing, we make a point of academic interest. The nature of 
the outer fixed point $\mathrm{C_2}$ in Fig.~\ref{f.1}, has a bearing 
on the possible range of values that the constant specific angular 
momentum $L$ may be allowed. For solutions passing through the outer
critical point, we can easily recognise from Eq.(\ref{critcon}) that 
the flow would be sub-Keplerian~\cite{az81,skc90}. In that case, we see 
that the condition $(L/L_{\mathrm{Kc2}})^2<1$ would hold good. In 
addition to this, from Eq.(\ref{eigen}), the saddle type behaviour 
of the point $\mathrm{C_2}$ would also imply that there would be 
another upper bound on $L$, given by $(L/L_{\mathrm{Kc2}})^2< \zeta$. 
For the admissible range of the polytropic index $\gamma$, the possible 
range for $\zeta$ would be $0< \zeta < 1/2$. This would then imply that 
the latter bound on $L$ would be more restrictive, as compared to the 
former. It is interesting that this essentially physical conclusion 
could be drawn from modelling the disc as a dynamical system in a 
mathematical parameter space. 

\section{A linearised perturbative analysis in real time}

As a preliminary study, the issue of explicit time-dependence in the 
accretion disc could be taken up from the viewpoint of a perturbative
analysis in real time. In the thin disc approximation, with the help of
a polytropic equation of state, the time-dependent generalisation of the 
radial momentum balance condition can be rendered as 
\begin{equation}
\label{tradmom}
\frac{\prt v}{\prt t} +v \frac{\prt v}{\prt R} 
+k \gamma  \rho ^{\gamma -2} \frac{\prt \rho}{\prt R} +
\frac{GM}{R^2} - \frac{L_{\mathrm{eff}}^2}{R^3} =0
\end{equation}
while the corresponding generalisation of the continuity equation can
be set down as $R{\prt_t}(\rho H) + (\rho vRH)^{\prime} = 0$, 
and the latter, 
in terms of a new variable defined as $f= \rho^{(\gamma +1)/2}vR^{5/2}$,
can be further recast as
\begin{equation}
\label{tconf}
\frac{\prt}{\prt t} \left[\rho^{(\gamma +1)/2} \right] +
\frac{1}{R^{5/2}} \frac{\prt f}{\prt R}=0
\end{equation}
The angular momentum balance condition may similarly be represented as 
\begin{equation}
\label{tangmom}
\frac{1}{v}\frac{\prt}{\prt t}(R^2 \Omega) + \frac{\prt}{\prt R}
(R^2 \Omega) = \alpha \left(\frac{\gamma k}{GM}\right)\frac{1}{f}
\frac{\prt}{\prt R}\left[f\left(\frac{f^2 \Omega_{\mathrm{K}}}
{\rho^2 v^3} \frac{\prt \Omega}{\prt R} \right)\right]
\end{equation}
From Eq.(\ref{tconf}) it is obvious that the steady solution of $f$,
given as $f_0$, would be a constant, which might, within a numerical
factor, be physically identified
with the steady mass inflow rate. For orbits in a fixed 
gravitational potential, the smallness of the change in angular velocity
allows us to ignore its explicit time variation~\cite{fkr92,pri81}. 
With the use of our quasi-viscous prescription, 
$R^2 \Omega = L + \alpha R^2 \Omega^{\prime}$,
in Eq.(\ref{tangmom}), we then get 
\begin{equation}
\label{tanginteg}
R^2 \Omega^{\prime} = -2L \left(\frac{\gamma k}{GM}\right) 
\left[ \frac{1}{R^3}
\frac{f^2 \Omega_{\mathrm{K}}}{\rho^2 v^3} + \int \frac{1}{R^3}
\frac{f^2 \Omega_{\mathrm{K}}}{\rho^2 v^3} \left(\frac{1}{f}
\frac{{\mathrm{d}}f}{{\mathrm{d}}R} \right) {\mathrm{d}} R \right]
+ L^{\prime}
\end{equation}
where $L^{\prime}$ is an integration constant. The effective angular
momentum is now defined as 
$L_{\mathrm{eff}} = L + \alpha R^2 \Omega^{\prime}$
and this is what we use in Eq.(\ref{tradmom}) after linearising in 
$\alpha$. It is easy to see that in the steady limit $\Omega^{\prime}$
should assume the form of $\tilde{\Omega}$ as given by Eq.(\ref{formofl})

The steady state solutions are given as $v_0$, $\rho_0$ and $f_0$, and
about these we impose small time-dependent perturbations $v^{\prime}$,
$\rho^{\prime}$ and $f^{\prime}$. In terms of $f^{\prime}$, we can
then derive an equation for the perturbation that is given by
\begin{align}
\label{tpert}
\frac{{\prt}^2 f^{\prime}}{\prt t^2} +2 \frac{\prt}{\prt R} 
\left(v_0 \frac{\prt f^{\prime}}{\prt t} \right) & +\frac{1}{v_0} 
\frac{\prt}{\prt R}\left[ v_0 \left(v_0^2- \beta^2 c_{\mathrm{s0}}^2 \right)
{\frac{\prt f^{\prime}}{\prt R}}\right] \nonumber \\
& -\frac{4 \alpha L^2}{R^3 v_0} \sigma \left[ \frac{1}{\sigma} 
\int \left(\frac{{\mathrm{d}} \sigma}{{\mathrm{d}} R} \right)
\frac{\prt f^{\prime}}{\prt t} {\mathrm{d}} R 
+ \left(\frac{3\gamma -1}{\gamma +1}\right) 
v_0 \frac{\prt f^{\prime}}{\prt R} \right] =0 
\end{align}
in which $\sigma = c_{\mathrm{s0}}^2/(v_0 v_{\mathrm{K}})$ and 
$\beta^2 = 2/(\gamma +1)$. 

We treat the perturbation as a standing wave and confine our analysis
to the outer subsonic region of the flow. From Fig.~\ref{f.2}, one can
see that in the vicinity of the focus the solutions would be 
double-valued. This is physically not feasible, and therefore for all
the solutions which spiral in towards the focus, their inner branch 
(for $R<R_\mathrm{c1}$) and their outer branch (for $R>R_\mathrm{c1}$)
would have to be fitted via a vertical discontinuity in the neighbourhood
of the focus. One may conceive of this discontinuity as a standing shock. 
For $R>R_\mathrm{c1}$, we will have a family of continuous 
solutions in the subsonic region, extending
from the shock to very great distances from the accretor. Somewhere in
this range, at two chosen points, we can constrain the perturbation
spatially by requiring the standing wave to die out at those points. 
The outer point could suitably be chosen to be at the outer boundary of 
the flow itself, where by virtue of the boundary condition on the steady 
flow, the perturbation would naturally decay out. We choose the inner 
point to be infinitesimally close to the shock front, which can 
behave like a fully absorbing wall for all disturbances
and make the perturbation vanish here too. Between these two points, the
time-dependent behaviour of the amplitude of the standing wave will be
an indicator of the stability of the steady solutions. For the 
perturbation we use a solution of the kind 
$f^{\prime}=p(R) \exp(-i \omega t)$. Subsequently we multiply the 
result that we get from Eq.(\ref{tpert}) by $v_0 g$, and then carry out
an integration by parts. This, followed by requiring that all the 
integrated ``surface" 
terms vanish at the two boundaries of the standing wave, will 
give us a quadratic in $\omega$, and it will be easy to see that 
$-i \omega$ will have a viscosity-dependent real part, even for 
arbitrarily small non-zero values of $\alpha$. This real part is 
given by 
\begin{equation}
\label{realpart}
\Re(-i\omega ) = 2\alpha \left(\int v_0 p^2{\mathrm{d}}R \right)^{-1}
\int v_0 p^2 \left[ \frac{L^2}{R^3 v_0 p} \int p \left( 
\frac{{\mathrm{d}} \sigma}{{\mathrm{d}} R} \right) {\mathrm{d}}R 
\right] {\mathrm{d}}R
\end{equation}
If $\Re(-i\omega ) > 0$, then the standing wave will have an amplitude
growing in time. The term in the square brackets above will be crucial in
this regard, and from its long-distance asymptotic behaviour at least, 
it is seen to be positive. Therefore on large length scales this 
indicates instability. 

To bolster our argument for this instability, we present the salient
features of an analysis in which the perturbation has been treated as a 
radially travelling wave. To that end we prescribe the spatial part of 
the perturbation as $p(R)=e^s$, where $s$ is given by the power series 
$s= \sum_{n=-1}^{\infty} k_n (R) \omega^{-n}$. We confine our analysis 
to the short wavelength regime, in which the wavelengths are considered
to be small compared to a relevant length scale in the system. 
The radius of the accretor, $R_\star$, could be chosen as such a 
limiting length scale. 

To make any further progress we need to examine the structure of the
function $s$. For that we go back to Eq.(\ref{tpert}) and use
the solution $f^{\prime}(R,t)= \exp(s - i\omega t)$ in this equation. 
In the resulting power series expansion, we sum up all 
the coefficients of the same power in $\omega$, and set each individual 
sum to zero. In that way all the coefficients involving $\omega^2$ will 
deliver the result 
\begin{equation}
\label{kayminus1}
k_{-1} = \int \frac{i}{v_0 \pm \beta c_{\mathrm{s0}}}{\mathrm{d}}R
\end{equation}
while the coefficients of $\omega$ will lead to 
\begin{equation}
\label{kaynot}
k_0 = - \frac{1}{2} \ln \left(\beta v_0 c_{\mathrm{s0}} \right)
\pm \alpha L^2 \left(3 \gamma -1 \right) \int 
\frac{\beta c_{\mathrm{s0}}\left(v_0 \pm \beta c_{\mathrm{s0}}\right)}
{v_0 v_{\mathrm{K}} R^3\left(v_0^2 - \beta^2 c_{\mathrm{s0}}^2\right)}
{\mathrm{d}}R
\end{equation}
The two expressions above give the leading terms in the power series 
of $s$. For self-consistency it will be necessary to show that 
successive terms follow the condition 
$\omega^{-n}\vert k_n(R)\vert \gg {\omega}^{-(n+1)}\vert k_{n+1}(R)\vert$.  
In the inviscid limit, this requirement can be shown to be very 
much true asymptotically for the first three terms, because we get 
$k_{-1} \sim R$, $k_0 \sim \ln R$ and $k_1 \sim R^{-1}$. With the 
inclusion of viscosity as a physical effect, we see from 
Eqs.(\ref{kayminus1}) and (\ref{kaynot}) respectively, that while 
$k_{-1}$ remains unaffected, $k_0$ acquires a viscosity-dependent
term that goes asymptotically as $\alpha R$. This in itself is an 
indication of the extent to which viscosity might alter the inviscid 
conditions. However, since we have chosen $\alpha$ to be very much 
less than unity, and since we have also chosen only short wavelength 
perturbations, implying that
$\omega \gg (v_0 \pm \beta c_{\mathrm{s0}})/ R_\star$,
our self-consistency requirement still holds. Therefore, in so far 
as we are looking for a qualitative understanding of the effect of 
viscosity, it should be sufficient for us to consider the two 
leading terms only in the power series expansion of $s$, and with 
the help of these two, one may then set down an expression for the 
perturbation as
\begin{equation}
\label{fpertur}
f^{\prime}(R,t) \simeq \frac{A_\pm}
{\sqrt{\beta v_0 c_{\mathrm{s0}}}}e^{-i \omega t}
\exp \left ( \int \left [ \frac{i \omega}{v_0 \pm \beta c_{\mathrm{s0}}}
\pm \alpha L^2 \left(3 \gamma -1 \right)
\frac{\beta c_{\mathrm{s0}}\left(v_0 \pm \beta c_{\mathrm{s0}}\right)}
{v_0 v_{\mathrm{K}} R^3\left(v_0^2 - \beta^2 c_{\mathrm{s0}}^2\right)}
\right]{\mathrm d}R \right)
\end{equation}
which should be seen as a linear superposition
of two waves with arbitrary constants $A_+$ and $A_-$. Both these
two waves move with a velocity $c_{\mathrm{s0}}$ relative to the fluid,
one against the bulk flow and the other along with it, while the bulk
flow itself has a velocity $v_0$. It should be immediately evident to us
that all questions pertaining to the growth or decay in the amplitude of
the perturbation will be crucially decided by the real terms delivered
to us from $k_0$. The viscosity-dependent term is especially influential
in this regard. For the choice of the lower sign in the real part of 
$f^\prime$ in Eq.(\ref{fpertur}), i.e. for the outgoing mode of the 
travelling wave solution, we see that the presence of viscosity causes 
the amplitude of the perturbation to diverge exponentially on large length 
scales (where $c_{\mathrm{s0}} \simeq c_{\mathrm{s}}(\infty)$ and 
$v_0 \sim R^{-5/2}$), since $-v_0$ is positive for inflows. 
The inwardly travelling mode also displays similar behaviour, albeit
to a quantitatively lesser degree. 
It is an easy exercise to see that stability in the system would be 
restored for the limit of $\alpha = 0$, and this particular issue has 
been discussed elsewhere~\cite{ray03}. The exponential growth behaviour
of the perturbation, therefore, is exclusively linked to the presence
of viscosity. 
Regarding this point, we may also mention that Kato et al.~\cite{kat88}
pointed out the existence of a non-propagating growing perturbation
localised at the critical point, but interestingly enough they also
found that this instability disappears in inviscid transonic flows. 
Chen and Taam~\cite{ct93} also made a numerical study of instability
in the disc. For short wavelength radial perturbations, they found
that the inertial-acoustic modes are locally unstable throughout
the disc, with the outward travelling modes growing faster than the
inward travelling modes in most regions of the disc, something that
is very much in keeping with what Eq.(\ref{fpertur}) indicates here. 

Since we are more concerned with the question of the preference of
the thin disc system for any particular solution, it now suffices
for us to appreciate that in this regard, a perturbative time-dependent 
analysis offers no clue. For the inviscid disc all physically
meaningful inflow solutions are stable, while with the introduction of
viscous dissipation these solutions are destabilised on large length
scales --- a region through which all inflows will have to pass anyway. 

In concluding this section it would be instructive to furnish a parallel 
instance of the destabilising influence of viscous dissipation in a system 
undergoing rotation : that of the effect of viscous dissipation in a 
Maclaurin spheroid~\cite{schan87}. 
In studying ellipsoidal figures of equilibrium, 
Chandrasekhar has discussed that secular instability (which manifests 
itself only if some dissipative mechanism is operative) arises in a
Maclaurin spheroid, when the stresses derive from an ordinary viscosity
which is defined in terms of a coefficient of kinematic viscosity (as
the $\alpha$ parametrisation of Shakura and Sunyaev is for an accretion 
disc), and when the effects arising from viscous dissipation are 
considered as small perturbations on the inviscid flow and taken 
into account in the first order~\cite{schan87}. 
It is exactly in this spirit that 
we have prescribed the ``quasi-viscous" approximation in the thin 
accretion disc.  

\section{Temporal evolution as a selection mechanism for transonicity}

In trying to make a time-dependent study, we have by now assured 
ourselves that a linearised perturbative analysis in real time cannot
give us any conclusive insight about the accreting system showing any 
preference for any particular solution. If at all the stability analysis 
has indicated anything worth a serious consideration, it is that the 
presence of viscosity tends to destabilise the thin disc system. One may 
then say that any selection mechanism based on explicit time-dependence 
has to be non-perturbative and evolutionary in character. We now proceed 
to make a study of that, confining our discussion to only the inviscid 
disc, because we have seen that the steady solutions of the inviscid 
model are stable under small perturbations, and we may therefore expect
stationary behaviour in the long-time limit. However, even for this 
simple inviscid system, with 
explicit time-dependence taken into consideration, the equations of a 
compressible flow cannot be integrated. Hence, to have any appreciation 
of the time-evolutionary selection of the critical solution, it should 
be instructive to consider an analogous model situation first. The model 
system being introduced here, describes the dynamics of the field 
$y(x,t)$ as
\begin{equation}
\label{mod}
\frac{\partial y}{\partial t}+\left(y-x\right)\frac{\partial y}{\partial x}
=y+2x+x^2
\end{equation}
whose static limit leads to 
\begin{equation}
\label{statmod}
\frac{\mathrm{d}y}{\mathrm{d}x} = \frac{y+2x+x^2}{y-x}
\end{equation}
and which, viewed as a dynamical system, is seen as 
\begin{align}
\label{dynmod}
\frac{\mathrm{d}y}{\mathrm{d}\tau} &= y+2x+x^2 \nonumber \\
{\frac{\mathrm{d}x}{\mathrm{d}\tau}} &= y-x
\end{align}

In the $y$ --- $x$ space, the fixed points 
$(x_{\mathrm c},y_{\mathrm c})$ are to be found at
$(0,0)$ and $(-3,-3)$. A linear stability
analysis of the fixed points in $\tau$ space gives the eigenvalues 
$\lambda$, by ${\lambda}^2 = 1+2(x_{\mathrm c} +1)$. It is then easy to 
see that $(0,0)$ is a saddle point while $(-3,-3)$ is a centre type point. 

\begin{figure}
\begin{center}
\includegraphics[scale=0.4, angle=-1.0]{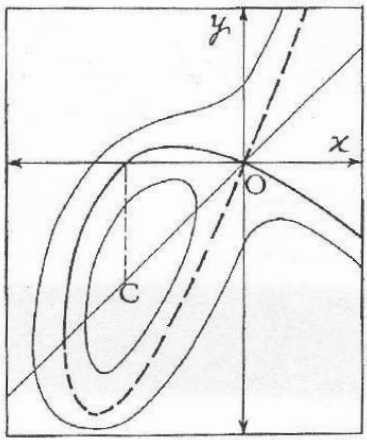}
\caption{\label{f.3} \small{A schematic diagram of the model problem. 
The curve given by Eq.(\ref{integmod}) passes through the origin 
for $c=0$. For this critical trajectory $\mathrm{O}$ is a saddle point 
while $\mathrm{C}$ is a centre type fixed point. This figure differs 
from Fig.~\ref{f.1} by a tilt of the axes.}}
\end{center}
\end{figure}

The integral curves are
\begin{equation}
\label{integmod}
y^2-2xy-2x^2- \frac{2}{3} x^3=c
\end{equation}
and the trajectories passing through the saddle point are the ones with
$c=0$. The different possible trajectories (drawn without arrows
associated with them) are shown in Fig.~\ref{f.3}. The separatrices are 
shown as the heavy solid curve and the heavy dashed curve. The similarity 
between Fig.~\ref{f.1} and Fig.~\ref{f.3} is pretty much in evidence. 

To explore the temporal dynamics, and to obtain a solution to 
Eq.(\ref{mod}), it would be necessary to apply the method of 
characteristics~\cite{deb97}. This involves writing
\begin{equation}
\label{charac}
{\frac{\mathrm{d}t}{1}}={\frac{\mathrm{d}x}{y-x}}= 
\frac{\mathrm{d}y}{y+2x+x^2}
\end{equation}

The task is to find two constants $c_1$ and $c_2$ from the above set and 
the general solution of Eq.(\ref{mod}) would then be given by ${c_1}
= F(c_2)$, where the function $F$ is to be determined from the initial
conditions. It is easy to see that one of the constants of integration
is clearly the $c$ of Eq.(\ref{integmod}). Hence, writing ${c_1}=c$ and 
on using Eq.(\ref{integmod}) in the first part of Eq.(\ref{charac}), 
we get
\begin{equation}
\label{charsol}
\int \, \mathrm{d}t= \pm \int \frac{\mathrm{d}x}
{\sqrt{3x^2+\left(2/3\right)x^3+c}}
\end{equation}
which solves the problem in principle. To put this in a usable form,
the integration in Eq.(\ref{charsol}) would have to be carried out. This 
cannot be done exactly. For small $x$ (the most important region, since it 
is near the saddle), the $x^3$ term may be left out to a good approximation. 
Further, only the positive sign in the right hand side of Eq.(\ref{charsol}) 
is to be chosen by the physical argument that the system is to evolve 
through a positive range of $t$ (time) values. Integration of
Eq.(\ref{charsol}) will then lead to the result
\begin{equation}
\label{timeinteg} 
\left( x+ \sqrt{x^2+c/3} \right) e^{- \sqrt{3} t}={c_2}
\end{equation}
which will then make the solution of Eq.(\ref{mod}) look like
\begin{equation}
\label{gensol}
y^2-2xy-2x^2- \frac{2}{3} x^3=F\left (\left [x+ 
\sqrt{ \frac{\left(y-x \right)^2}{3}
- \frac{2}{9}x^3}\right ]e^{-{\sqrt 3}t} \right )
\end{equation}

The evolution of the system is to be followed from the initial 
condition that $y=0$ for all $x$ at $t=0$. Dropping the $x^3$ term again
for small $x$, allows us to determine the form of the function 
$F$ as $F(z)=-3(2- \sqrt{3})z^2$. The solution, consequently, becomes 
\begin{equation}
\label{solinit}
y^2-2xy-2x^2- \frac{2}{3} x^3=-3\left(2- \sqrt{3}\right)
\left [ x+ \sqrt{ \frac{\left(y-x \right)^2}{3}-
\frac{2}{9}x^3}\right ]^2 e^{-2{\sqrt{3}}t}
\end{equation}
and as $t{\longrightarrow}{\infty}$, the steady solution that 
would be selected would be 
\begin{equation}
\label{integmod2}
y^2-2xy-2x^2- \frac{2}{3} x^3=0
\end{equation}
which is actually the equation for the separatrices. 
It is worth stressing the remarkable feature of this result. The
evolution started under conditions far removed from transonicity. In 
fact, it started as $y=0$ for all $x$ (in the vicinity of the origin of 
coordinates) at $t=0$. The evolution proceeded through a myriad of 
possible steady state solutions (all arguably stable under a linear 
stability analysis) and then in the stationary limit, selected the 
separatrices. This is a convincing demonstration that it is in principle 
possible for apparently non-realisable separatrices in the steady regime, 
to become eminently realisable physically, when the temporal evolution of 
the system is followed. 

To the extent that the model problem has been a good representative of the 
true physical situation, the whole argument 
for a time-dependent and non-perturbative method of selection developed 
above, may now be extended to the actual problem of thin disc accretion.
For a steady accretion disc many previous works have upheld the case for 
transonicity~\cite{lt80,skc90}, although without explicitly addressing the 
issue of what special physical criterion may select the transonic solution 
to the exclusion of all other possible solutions. For the case of disc 
accretion on to black holes, Liang and Thompson~\cite{lt80} make a 
clear point by saying that ``the solution for the radial drift velocity of 
thin disk accretion onto black holes must be transonic, and is analogous to 
the critical solution in spherical Bondi accretion, except for the presence 
of angular momentum". For the inviscid disc at least, this is a most crucial 
analogy. In our chosen disc model, of course, the gravitational 
attractor driving the accretion process is not a black hole, but the 
question of the transonicity of the inflow solution is not expected to be 
too radically affected by this difference, especially in the vicinity of 
the far-off outer critical (saddle) point. However, what is important for 
us is the similarity with spherical symmetry that 
Liang and Thompson~\cite{lt80} have indicated. The inviscid and thin disc 
model has rendered the related mathematical problem very conveniently 
one-dimensional, and has made the analysis very similar to a spherically 
symmetric flow. From this analogy it should be instructive
for us at this point to derive some useful physical insight. 

Transonicity is a settled fact~\cite{bon52,gar79} in spherically 
symmetric accretion, and this has been so because at
every spatial point in the flow, velocity evolves at a much greater
rate through time, in comparison with density, and this therefore excludes
the possibility of matter accumulating in regions close to the surface of
the accretor. As a result gravity wins over pressure at small distances
and the system is naturally driven towards selecting the transonic
solution. This then raises the question of whether or not a likewise 
time-evolutionary mechanism should be at work for the selection of the 
critical inflow solution in an inviscid and thin accretion disc. 
We contend that it should be so. 

To have any appreciation of how the temporal evolution acts as a selection 
mechanism in disc accretion, it should be necessary to consider the dynamic 
equation for velocity evolution in the inviscid and thin accretion 
flow. This is given by
\begin{equation}
\label{timdep}
{\frac{\partial v}{\partial t}}+v{\frac{\partial v}{\partial R}}
+k{\gamma}{\rho}^{\gamma -2}{\frac{\partial \rho}{\partial R}}+
{\frac{GM}{R^2}} -{\frac{L^2}{R^3}} = 0
\end{equation}
It would be easy to see that to have a solution pass through the 
outer critical point (the saddle point), the temporal evolution should 
proceed in such a manner, that the inflow velocity would increase much 
faster in time than density
at distances given by $R \sim R_{c2}$, where for sub-Keplerian 
flows, gravity, going as $R^{-2}$, dominates the rotational effects,
which go as $R^{-3}$. That the velocity starts decreasing in the region 
where $R< R_{c1}$, is because of the resistance arising due to centrifugal 
effects in the fluid flow, which is quite unrelated to the density. 
The role of the pressure term (deriving from density) to resist the 
flow is confined to the subcritical solutions only. 
 
But what should be the key physical criterion guiding this dynamic 
selection of the transonic solution? We stress once again that the 
selection principle would very likely be the same as in the spherically 
symmetric case, where the transonic solution is chosen by dint of its 
corresponding to a configuration of the lowest possible 
energy~\cite{gar79}. To have an understanding of this, we consider 
Eq.(\ref{timdep}), with the density gradient being neglected as a  
working approximation. On large length scales, where the flow is 
highly subsonic, this is an especially effective approximation, and
it allows us to treat the evolution of the velocity field to be 
largely independent of the density evolution. We then integrate
the rest of the partial differential equation for velocity by the method 
of characteristics~\cite{deb97} and get a solution that can be written 
down as 
\begin{equation}
\label{emin}
\frac{v^2}{2} - \frac{GM}{R} + \frac{L^2}{2R^2} = \tilde{F}
\left[ \frac{1}{R_0 + R\left(1 + v/C \right)} \exp \left( \frac{vR}{CR_0} 
- \frac{Ct}{R_0} \right) \right]
\end{equation}
in which $C$ is an integration constant and $R_0 = GM/C^2$. The form 
of the function $\tilde{F}$ is to be determined from the physically 
realistic initial condition $v=0$ at $t=0$ for all $R$, which will 
render $\tilde{F}$ as 
\begin{equation}
\label{eff}
\tilde{F}(\xi) = - \frac{GM \xi}{1 - \xi R_0} + \frac{L^2 {\xi}^2}
{2\left(1 - \xi R_0 \right)^2} 
\end{equation}
If we examine the argument of $\tilde{F}$ in Eq.(\ref{emin}) and then 
study its long-time behaviour, we will see from Eq.(\ref{eff}) that for
$t \longrightarrow \infty$, the selected solution will correspond to
\begin{equation}
\label{select}
\frac{v^2}{2} - \frac{GM}{R} + \frac{L^2}{2R^2} = 0
\end{equation}

We can now see that prior to the evolution, the system had no bulk 
motion and that the radial drift velocity was given flatly everywhere 
by $v=0$. This, of course, gives the condition that initially the total
specific mechanical energy of the system was zero. Then at $t=0$ both 
a gravitational mechanism is activated in this system and some angular 
momentum is imparted to it. This will induce a potential $-GM/R$ 
everywhere, and at the same time start a rotational 
motion, respectively. The system will then 
start evolving in time, with the velocity $v$ at each point in space
evolving temporally according to Eq.(\ref{emin}). Finally the system will 
restore itself to a steady state in such a manner that the total 
specific mechanical energy at the end of the evolution 
(for $t \longrightarrow \infty$) 
will remain the same as at the beginning (at $t=0$), a condition that is 
given by Eq.(\ref{select}), whose left hand side gives the sum of the
specific kinetic energy, the specific gravitational potential energy 
and the specific rotational energy. 
This sum is zero, and therefore, under the given initial condition,
this must be the steady state corresponding to the minimum possible total 
specific energy of the system. Hence, this is the configuration that
is dynamically and non-perturbatively selected. It is now conceivable 
that if the pressure term were to be taken into account, then the 
solution that would be dynamically selected, would be the one that 
would pass through the saddle point, since, as Bondi had analogously 
conjectured for spherically symmetric accretion~\cite{bon52}, this 
would be the one to satisfy the criterion of minimum energy. 

Interestingly enough, this question can also be addressed from a very
different perspective. We have already derived an equation for
a perturbation in the flow, as given by Eq.(\ref{tpert}). For the 
inviscid disc that we are studying, an intermediate step in arriving 
at this expression, with $\alpha = 0$, is 
\begin{equation}
\label{interm}
\frac{\prt}{\prt t} \left[\frac{v_0}{f_0}
\left( \frac{\prt f^{\prime}}{\prt t}\right)\right]
+ \frac{\prt}{\prt t} \left[\frac{v_0^2}{f_0}
\left( \frac{\prt f^{\prime}}{\prt R}\right)\right]
+ \frac{\prt}{\prt R} \left[\frac{v_0^2}{f_0}
\left( \frac{\prt f^{\prime}}{\prt t}\right)\right] 
+ \frac{\prt}{\prt R} \left[\frac{v_0}{f_0}
\left(v_0^2 - \beta^2 c_{\mathrm{s0}}^2 \right) 
\frac{\prt f^{\prime}}{\prt R}\right] = 0 
\end{equation}
all of whose terms can be ultimately rendered into a compact 
formulation that looks like~\cite{vis98}
\begin{equation}
\label{compact}
\prt_\mu \left( {\mathrm{f}}^{\mu \nu} \prt_\nu
f^{\prime}\right) = 0
\end{equation}
in which we make the Greek indices run from $0$ to $1$, with the
identification that $0$ stands for $t$, and $1$ stands for $R$.
An inspection of the terms in the left hand side of Eq.(\ref{interm})
will then enable us to construct the symmetric matrix
\begin{equation}
\label{matrix}
{\mathrm{f}}^{\mu \nu } = \frac{v_0}{f_0}
\begin{bmatrix} 
1 & v_0 \\
v_0 & v_0^2 - \beta^2 c_{\mathrm{s0}}^2
\end{bmatrix} \\
\end{equation}
Now the d'Alembertian for a scalar in curved space is given in terms 
of the metric ${\mathrm{g}}_{\mu \nu}$ by~\cite{vis98} 
\begin{equation}
\label{alem}
\Delta \psi \equiv \frac{1}{\sqrt{-\mathrm{g}}}
\prt_\mu \left({\sqrt{-\mathrm{g}}}\, {\mathrm{g}}^{\mu \nu} \prt_\nu
\psi \right)
\end{equation}
with $\mathrm{g}^{\mu \nu}$ being the inverse of the matrix implied 
by ${\mathrm{g}}_{\mu \nu}$. We use the equivalence that 
${\mathrm{f}}^{\mu \nu } = \sqrt{-\mathrm{g}}\, {\mathrm{g}}^{\mu \nu}$, 
and therefore $\mathrm{g} = \det \left({\mathrm{f}}^{\mu \nu }\right)$, 
to immediately set down an effective metric for the propagation of an
acoustic disturbance as
\begin{equation}
\label{metric}
\mathrm{g}^{\mu \nu}_{\mathrm{eff}} = 
\begin{bmatrix}  
1 & v_0 \\
v_0 & v_0^2 - \beta^2 c_{\mathrm{s0}}^2
\end{bmatrix} \\
\end{equation}
which can be shown to be entirely identical to 
the metric of a wave equation for a
scalar field in curved space-time~\cite{vis98}. The inverse effective 
metric, $\mathrm{g}_{\mu \nu}^{\mathrm{eff}}$, can be easily obtained 
from Eq.(\ref{metric}), and this will give 
$v_0^2 = \beta^2 c_{\mathrm{s0}}^2$ as the horizon condition of an 
acoustic black hole~\cite{vis98}. Exploiting this close correspondence
between the physics of supersonic acoustic flows and many features of 
black hole physics, we can argue that since infalling matter crosses the 
event horizon of a black hole maximally, i.e. at the greatest possible 
speed, by analogy the same thing may be said of the manner in which a
rotating flow in an inviscid disc crosses an analogous acoustic horizon,  
given by the 
critical condition 
$v_0^2 = \beta^2 c_{\mathrm{s0}}^2$. This could only mean that the 
flow will cross the acoustic horizon transonically, since it is with
the transonic flow that the greatest possible flow rate is associated. 
That we should be able to appreciate this fact through what is in 
essence a perturbative method, as given by Eq.(\ref{interm}), is quite 
remarkable, because conventional wisdom tells us that we should be quite 
unable to have any understanding of the special status of any inflow 
solution solely through a perturbative technique~\cite{gar79}.  

Having gained this insight into the preference for the transonic 
solution in an inviscid disc, it should also be quite obvious for us 
to realise that the introduction of any viscosity-dependent term in 
Eq.(\ref{interm}) will disrupt the precise symmetry of Eq.(\ref{compact}). 
With this disruption of the symmetry obtained from inviscid conditions,
we should also lose all clear-cut vision of the way in which matter will
cross the acoustic horizon. That this is what it should be in the presence
of dissipation in the system, could also be understood from other
considerations. The  
quasi-viscous disc being a dissipative system, i.e. energy being allowed
to be drained away from this system, there should be no occasion for
us to look for the selection of a particular solution, and a selection
criterion thereof, on the basis of energy minimisation, as we have 
been able to conjecture for the idealised inviscid disc. This, of
course, shall also affect the flow rate, and the whole system would
be left without any well-defined criterion by which it could guide 
itself towards a steady state. Secondly,
already having acquainted ourselves with the fact that the quasi-viscous
disc is unstable on large length scales, we should expect no solution
--- transonic or otherwise --- to be free of time-dependence. Therefore,
a long-time evolution of the disc towards a stationary end is not 
something that we might hope for.  

Nevertheless, we are at least well aware of the fact that viscosity
determines the distribution of matter in a viscous disc. This, of course,
is intimately connected with the cumulative transfer of angular 
momentum to large length scales of the disc. If we refer to 
Eq.(\ref{asympangmom}), we shall see that in the outer regions of 
the disc, the effective specific force is given to a first order
in $\alpha$ by 
\begin{equation}
\label{effforce}
f_{\mathrm{eff}}(R) \sim - \frac{GM}{R^2} + \frac{L^2}{R^3}
+ 4\alpha \frac{L^2}{R_L^3} 
\end{equation}
from which it is evident that on scales of $R \sim R_L$, the transport
of angular momentum will give rise to an asymptotic constant non-zero 
force opposed to gravity. For the viscous disc this gives rise to 
a repulsive effect on large length scales vis-a-vis gravitational 
attraction. From this argument we go a step beyond and conjecture that 
$\alpha^{-1/3} R_L$ defines a limiting length scale for accretion 
that the outward transport of angular momentum imposes. This 
actually should be quite compatible with how viscosity redistributes 
an annulus of matter in a Keplerian flow around an accretor : 
the inner region drifting in because of dissipation, and consequently,  
through the conservation of angular momentum and its outward transport, 
making it necessary for the outer regions of the matter distribution 
to spread even further outwards~\cite{pri81,fkr92}. This state of 
affairs is qualitatively not altered in anyway for the quasi-viscous 
flow, except 
for the fact that with viscosity being very weak here, the outward 
transport of angular momentum can perceptibly cause an outward drift
of matter only on very large scales. It is obvious that once 
$\alpha =0$, i.e. for the inviscid limit, this scale would be shifted
to infinity. 

\begin{acknowledgments}
This research has made use of NASA's Astrophysics Data System.
One of the authors (AKR) gratefully acknowledges the support provided 
by the Council of Scientific and Industrial Research, Government of 
India, for a part of the time that was needed to carry out this work.
The authors would also like to thank Dr. Tapas Kumar Das, Prof. Rajaram 
Nityananda and Prof. Paul J. Wiita for some helpful comments. 
\end{acknowledgments}

\bibliography{archiv_disc}

\end{document}